\documentclass[twoside]{article}
\usepackage{fleqn,espcrc2}
\usepackage{amssymb}
\usepackage{epsfig}

\usepackage{graphicx}
\usepackage[figuresright]{rotating}

\newcommand{\AmS}{{\protect\the\textfont2
  A\kern-.1667em\lower.5ex\hbox{M}\kern-.125emS}}

\title{Four-neutrino mixing solutions 
		of the atmospheric neutrino anomaly}

\author{A. Marrone\address{Dipartimento di Fisica and Sezione INFN di Bari\\
             	Via Amendola 173, I-70126 Bari, Italy }}
       
\begin{document}

\begin{abstract}
Solutions to the atmospheric neutrino anomaly which smoothly
interpolate between $\nu_\mu\to\nu_\tau$ and $\nu_\mu\to\nu_s$ oscillations
are studied. It is shown
that, although the Super-Kamiokande data disfavor the {\em pure\/}
$\nu_\mu\to\nu_s$ channel,  one cannot exclude sizable
amplitude for the $\nu_\mu\to\nu_s$ channel {\em in addition} to 
$\nu_\mu \to \nu_\tau$ oscillations.
\vspace{1pc}
\end{abstract}

\maketitle

%%%%%%%%%%%%%%%%%%%%%%%%%%%%%%%%%%%%%%%%%%%%%%%%%%%%%%%%%%%%%%%%%%%%%%%%%%%%
\section{Introduction}
%%%%%%%%%%%%%%%%%%%%%%%%%%%%%%%%%%%%%%%%%%%%%%%%%%%%%%%%%%%%%%%%%%%%%%%%%%%%

Four-neutrino ($4\nu$)~\cite{Bi99,Do00,Ba00} models including
a hypothetical sterile 
state ($\nu_s$) can accommodate the three sources of evidence for $\nu$
flavor oscillations
coming from atmospheric, solar, and accelerator neutrino experiments. In
particular, $4\nu$ spectra with mass eigenstates  organized in two doublets
(2+2 models) seem to be favored by world  neutrino data \cite{Four}.
Although 2+2 models are often assumed to imply
the $\nu_\mu \to \nu_\tau$ or
$\nu_\mu \to \nu_s$ channel
for solar and atmospheric oscillations,  
one might have mixed (active+sterile) flavor
transitions of the kind \cite{Li2K}
%....................................................................................
\begin{eqnarray}
\nu_\mu &\to& \nu_+\;\; ({\rm atmospheric})\ ,\label{mu+}\\
\nu_e &\to& \nu_-\;\; ({\rm solar})\ ,\label{e-}
\end{eqnarray}
%....................................................................................
where  the states $\nu_\pm$, as discussed in detail in~\cite{Fo00},
represent
linear (orthogonal)  combinations of $\nu_\tau$ and $\nu_s$ through a mixing
angle $\xi$.
%\begin{eqnarray}
%\nu_+ &=& +\cos\xi\,\nu_\tau + \sin\xi\,\nu_s\ ,\label{nu+}\\
%\nu_-&=& -\sin\xi\, \nu_\tau + \cos\xi\, \nu_s\ .
%\label{nu-}
%\end{eqnarray}
%.............................................................................
The oscillation modes~(\ref{mu+},\ref{e-}) represent generalizations of both
modes $\nu_\mu\to\nu_\tau$ and $\nu_\mu\to\nu_s$, to which they reduce
for $\sin\xi=0$ and 1. For generic values of $\sin\xi$, the
final  states in atmospheric $\nu_\mu$ and solar $\nu_e$ flavor transitions are
linear combinations of $\nu_s$ and $\nu_\tau$, and the coefficients have to be
constrained by experiments.
In this work, four-neutrino oscillations in
the context of 2+2 spectra are studied, for unconstrained values of
$\sin\xi=\langle\nu_+|\nu_s\rangle$. It is shown that the state $\nu_+$ (into
which $\nu_\mu$ oscillates) can have a sizable $\nu_s$ component.
As shown in~\cite{Fo00},
these results
can be reconcilied to the $4\nu$ solutions to the
solar neutrino problem \cite{Pena,Pe2K}, which are compatible with a large
$\nu_s$ component of $\nu_-$.

%%%%%%%%%%%%%%%%%%%%%%%%%%%%%%%%%%%%%%%%%%%%%%%%%%%%%%%%%%%%%%%%%%%%%%%%%%%%
\section{Graphical representations}
%%%%%%%%%%%%%%%%%%%%%%%%%%%%%%%%%%%%%%%%%%%%%%%%%%%%%%%%%%%%%%%%%%%%%%%%%%%%
The mixing parameter spaces for
atmospheric and solar neutrinos can be represented in triangular
plots, embedding unitarity relations of the kind $U^2_1+U^2_2+U^2_3=1$,
%---------------
\footnote{Triangle plots have been already introduced and discussed in detail
in the context of $3\nu$ mixing, see \protect\cite{F3nu}.}
%-------------
holding for the four columns of the mixing matrix $U$,
under suitable approximations~\cite{Fo00}. 
It can be shown that is sufficient to implement  just  one unitarity
relation in one triangle plot for  the two mass doublets,
the first ($\nu_3$, $\nu_4$) being the atmospheric doublet, 
the second ($\nu_1$, $\nu_2$) the solar doublet, in our framework.
Under the semplifying hypotheses discussed in \cite{Fo00},
the $\nu_e$ does not take part to the oscillations of atmospheric neutrinos, so
that the mixing can be graphically represented in
a triangle plot, where the upper, lower left and lower right corner are
identified with $\nu_\mu$, $\nu_s$, and $\nu_\tau$, respectively,
and the actual mixing is determined by the position of the $\nu_4$ mass
eigenstate in the triangle.
The heights
projected from a generic point $\nu_4$ inside the triangle onto the lower,
right, and left side  represent the elements $U^2_{\mu 4}$, $U^2_{s 4}$, and
$U^2_{\tau 4}$, respectively. These three matrix elements can be expressed
as functions of two
mixing angles $(\psi\ , \xi)$. The square mass difference
$m^2$ between the
atmospheric doublet ($\nu_3$, $\nu_4$) completes the parametrization.
  When $\nu_4$ coincides with one of the corners 
no oscillation occur.  Generic
points inside the triangle
describe mixed (active+sterile) atmospheric neutrino oscillations, smoothly
interpolating from pure $\nu_\mu\to s$ to pure $\nu_\mu\to\nu_\tau$.

%%%%%%%%%%%%%%%%%%%%%%%%%%%%%%%%%%%%%%%%%%%%%%%%%%%%%%%%%%%%%%%%%%%%%%%%%%%%
\section{Results of the atmospheric $\nu$ analysis}
%%%%%%%%%%%%%%%%%%%%%%%%%%%%%%%%%%%%%%%%%%%%%%%%%%%%%%%%%%%%%%%%%%%%%%%%%%%%
\begin{figure}[t]
\mbox{
\psfig{bbllx=3.truecm,bblly=11.0truecm,bburx=10.0truecm,bbury=20.5truecm,
height=4.3truecm,figure=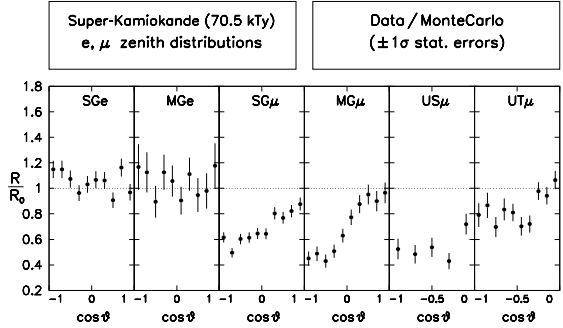}}
\caption{Super-Kamiokande data on zenith distributions of lepton events induced
by atmospheric neutrinos.
}
\label{nbest-dist}
\end{figure}
The most recent 70.5 kTy Super-Kamiokande 
data used in our analysis \cite{So2K} are shown in Figure~1.
The statistical $\chi^2$ analysis is similar to that of~\cite{F3nu}.
\begin{figure}[t]
\mbox{
\psfig{bbllx=1.5truecm,bblly=7.0truecm,bburx=4.0truecm,bbury=28.truecm,
height=8.8truecm,figure=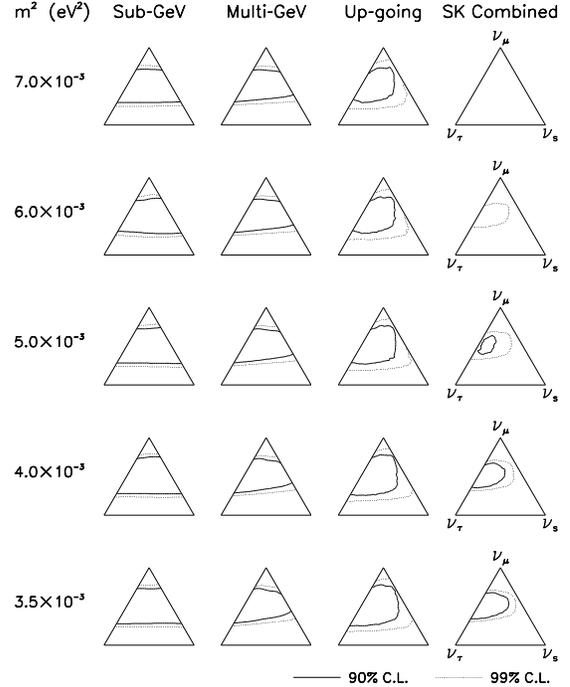}}
\caption{Results of the global fit to \mbox{atmospheric} neutrino data in the
triangular plots, for decresing $m^2$ values.
}
\label{nbest-dist}
\end{figure}
Figures~2 and 3 show the results of our analysis in the  atmospheric triangle
plot. The first three
columns of triangles refer to the separate fits  to sub-GeV electrons and muons
(10+10 bins), multi-GeV electrons and muons (10+10 bins), and upward stopping
and through-going muons (5+10 bins), while the fourth column refers to the
total SK data sample (55 bins). For each column, we find the minimum $\chi^2$
from the fit to the corresponding data sample, and then present sections of the
allowed volume at fixed values of $m^2$ for  $\Delta\chi^2=6.25$ (90\% C.L.,
solid lines) and $\Delta\chi^2=11.36$ (99\% C.L., dotted lines). 
\begin{figure}[t]
\mbox{
\psfig{bbllx=1.5truecm,bblly=7.0truecm,bburx=4.0truecm,bbury=28.truecm,
height=8.8truecm,figure=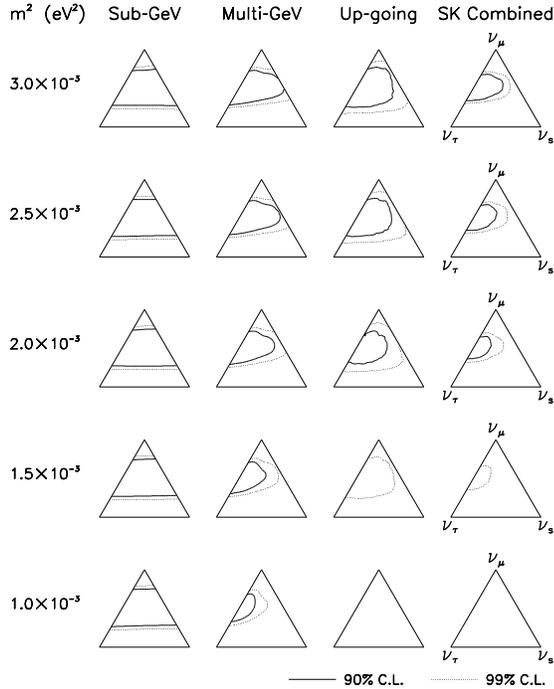}}
\caption{As in Fig. 2, for lower $m^2$ values.
}
\label{nbest-dist}
\end{figure}
The low-energy SG data are basically insensitive to $s^2_\xi$, since they
are consistent both with $\nu_\mu\to\nu_\tau$ oscillations (left side) and with
$\nu_\mu\to\nu_s$ oscillations (right side), as well as with any intermediate
combination of the two  oscillation channels. High-energy upgoing muon data are
instead much more sensitive to the $\nu_s$ component through matter effects,
which increase both with energy and with $s^2_\xi$, and tend to suppress the
muon deficit. Large matter effects appear to be  disfavored by the
upgoing muon data, the rightmost part of the triangle being excluded at 90\%
C.L. The multi-GeV data cover an intermediate energy range and are not as
constraining as the upgoing muons. However, they show a tendency to
disfavor a large sterile component, that is strengthened in the 
global combination of data (fourth column), leading to the upper bound
%....................................................................
\begin{equation}
s^2_\xi \lesssim 0.67\  (90\% {\rm\ C.L.})\ ,
\label{xibound}
\end{equation}
%...................................................................
which disfavors pure or quasi-pure $\nu_\mu\to\nu_s$ oscillations.
From Figs.~2 and 3 we derive that pure $\nu_\mu\to\nu_s$ 
oscillations
are excluded at $>99\%$ C.L., in agreement whith SK results~\cite{SK2K}.
However, the bound (\ref{xibound}) still allows intermediate cases of
oscillations 
with a sizable $\nu_s$ component. For instance,
one cannot exclude that the sterile neutrino channel may have the same
amplitude as the active one $(s^2_\xi=0.5)$, or may even be dominant.
Furthermore, for $m^2$ in its upper range,
a subdominant $\nu_s$ component actually helps in the global fit to
SK data, leading to a 90\% C.L. allowed region which does not touch the left
side of the triangle (pure $\nu_\mu\to\nu_\tau$ channel).

Figures~2 and 3 also provide bounds on the  other mixing parameter $s^2_\psi$,
which governs the relative amount of $\nu_\mu$ and $\nu_+$ on the atmospheric
neutrino states $\nu_3$ and $\nu_4$ 
and thus the overall amplitude of $\nu_\mu\to\nu_+$ oscillations.
The following limit can be derived
%....................................................................
\begin{equation}
s^2_\psi \simeq 0.51\pm 0.17\  (90\% {\rm\ C.L.})\ ,
\end{equation}
%...................................................................
which favors $\nu_\mu\to\nu_+$ oscillations with nearly maximal amplitude, as
expected from the observation of nearly maximal average suppression ($\sim
50\%$) of upgoing MG muons. 
Therefore, on the basis of present SK data on the
zenith distributions of leptons induced by atmospheric neutrinos, pure
$\nu_\mu\to\nu_s$ oscillations are strongly disfavored, but one cannot exclude
mixed active-sterile $\nu_\mu\to\nu_+$ oscillations, 
provided that the partial amplitude of
the sterile channel is $\lesssim 67\%$. 

%%%%%%%%%%%%%%%%%%%%%%%%%%%%%%%%%%%%%%%%%%%%%%%%%%%%%%%%%%%%%%%%%%%%%%%%%%%%
\section{Summary and conclusions}
%%%%%%%%%%%%%%%%%%%%%%%%%%%%%%%%%%%%%%%%%%%%%%%%%%%%%%%%%%%%%%%%%%%%%%%%%%%%

In the context of 2+2 neutrino models, the SK
atmospheric data 
have been studied, assuming the
coexistence of both $\nu_\mu\to\nu_\tau$ and $\nu_\mu\to\nu_s$ oscillations,
with a smooth interpolation between such two subcases. It has been shown that,
although the data disfavor oscillations in the pure $\nu_\mu\to\nu_s$ channel,
one cannot exclude their presence {\em in addition\/} to $\nu_\mu\to\nu_\tau$  
oscillations. High energy muon data appear to be crucial to determine
the relative
amplitude of the active and sterile oscillation channels for atmospheric
neutrinos.

%%%%%%%%%%%%%%%%%%%%%%%%%%%%%%%%%%%%%%%%%%%%%%%%%%%%%%%%%%%%%%%%%%%%%%%%%%%
% 			R E F E R E N C E S 
%%%%%%%%%%%%%%%%%%%%%%%%%%%%%%%%%%%%%%%%%%%%%%%%%%%%%%%%%%%%%%%%%%%%%%%%%%%%%%%


\begin{thebibliography}{45}
		
\bibitem{Bi99}	S.M.\ Bilenky, C.\ Giunti, and W.\ Grimus,
		Prog.\ Part.\ Nucl.\ Phys.\ {\bf 43}, 1 (2000).
		
\bibitem{Do00}	D.\ Dooling, C.\ Giunti, K.\ Kang, and C.W.\ Kim,
		Phys.\ Rev.\ D {\bf 61}, 073011 (2000).	
		
\bibitem{Ba00}	V.\ Barger and K.\ Whisnant, hep-ph/0006235,
		in {\em Current Aspects of Neutrino Physics},
		ed.\ by D.\ Caldwell (Springer-Verlag, Hamburg, 2000).	

\bibitem{Four}	S.M.\ Bilenky, C.\ Giunti, and W.\ Grimus,
		Proceedings of {\em Neutrino '96\/}, Helsinki, June 1996,
		edited by K.\ Enqvist {\em et al.} 
		(World Scientific, Singapore, 1997), p.174, hep-ph/9609343;
		V.\ Barger, S.\ Pakvasa, T.J.\ Weiler, and K.\ Whisnant,
		Phys.\ Rev.\ D {\bf 58}, 093016 (1998).

\bibitem{Li2K}	G.L.\ Fogli, E.\ Lisi, and A.\ Marrone,   
		in {\em Neutrino 2000}, 
		19th International Conference on Neutrino 
		Physics and Astrophysics (Sudbury, Canada, 2000),
		to appear; transparencies available at the site 
		http://nu2000.sno.laurentian.ca.	
		

\bibitem{Fo00}	G.L.\ Fogli, E.\ Lisi, and A.\ Marrone,   
		hep-ph/0009299, accepted by Phys.\ Rev.\ D.	
		
\bibitem{Pena}	C.\ Giunti, M.C.\ Gonzalez-Garcia, and 	C.\ Pe{\~n}a-Garay,
		Phys.\ Rev.\ D {\bf 62}, 013005 (2000). 
		
\bibitem{Pe2K}	M.C.\ Gonzalez-Garcia and C.\ Pe{\~n}a-Garay,
		hep-ph/0009041.
		
\bibitem{So2K}	H.\ Sobel for the Super-Kamiokande Collaboration,
		in {\em Neutrino 2000\/} \protect\cite{Li2K}.
		
\bibitem{F3nu}	G.L.\ Fogli, E.\ Lisi, A.\ Marrone, and G.\ Scioscia,
		Phys.\ Rev.\ D {\bf 59}, 033001 (1999).


		
\bibitem{SK2K}	Super-Kamiokande Collaboration, S.\ Fukuda {\em et al.},
		Phys.\ Rev.\ Lett. {\bf 85}, 3999 (2000).

\end{thebibliography}
\end{document}